\documentclass[a4paper]{article}

\usepackage{INTERSPEECH2020}
\usepackage{cite,multirow,balance}

\usepackage[colorlinks = true,
            linkcolor = blue,
            urlcolor  = blue,
            citecolor = blue,
            anchorcolor = blue]{hyperref}

\title{The Attacker's Perspective on Automatic Speaker Verification: An Overview}
\name{Rohan Kumar Das$^1$, Xiaohai Tian$^1$, Tomi Kinnunen$^2$ and Haizhou Li$^1$}
\address{
  $^1$Department of Electrical and Computer Engineering, National University of Singapore, Singapore\\
  $^2$School of Computing, University of Eastern Finland, Joensuu, Finland
  }
\email{\{rohankd, eletia, haizhou.li\}@nus.edu.sg, tkinnu@cs.uef.fi}

\begin{document}

\maketitle
\begin{abstract}
Security of automatic speaker verification (ASV) systems is compromised by various spoofing attacks. While many types of \emph{non-proactive} attacks (and their defenses) have been studied in the past,
\emph{attacker's} perspective on ASV, 
represents a far less explored direction. It can potentially
help to identify the weakest parts of ASV systems and be used to develop attacker-aware systems. 
We present an overview on this emerging research area by focusing on 
potential threats of adversarial attacks on ASV, spoofing countermeasures, or both. We conclude the study with discussion on selected attacks and leveraging from such knowledge to improve defense mechanisms against adversarial attacks. 
\end{abstract}

\vspace{2mm}

\section{Introduction}

\emph{Automatic speaker verification} (ASV) technology is now a matured technology used in access control, forensics and surveillance applications~\cite{Tomi,sv_debut}. Unfortunately, unprotected ASV systems are highly vulnerable to various \emph{spoofing attacks}~\cite{ISO_spoofing} where an attacker (adversary) masquerades him/herself as a specific targeted user. This has motivated the study of automatic detection of spoofing attacks~\cite{spoof_review}. 
Such \emph{countermeasures} have been studied as one of the important topics in system implementation, either independently of, or in conjunction with ASV.

\emph{ASVspoof} challenge series~\cite{ASVspoof_journal} is a community-driven bechmarking effort to address voice spoofing attacks and their defenses. 
The attacks include various \emph{voice conversion} (VC) and  \emph{text-to-speech synthesis} (TTS) techniques along with audio replay~\cite{ASVspoof2019_paper}. Their impact upon ASV is now far better understood than a decade ago. Nonetheless, vast majority of research in this domain focuses on \emph{non-proactive} attacks, where the adversary takes no direct use of the attacked system. For instance, the typical objective of VC and TTS is to maximize perceptual speaker similarity and audio quality, rather to break ASV systems. 

Apart from studying robust spoofing countermeasures, 
it is important to study the weak links of ASV to protect it from various types of attacks. In order to identify the loopholes of ASV, we need to assess the limits of spoofing attacks \emph{from the perspective of the attacker}. For an attacker, the ideal way is to attack within the functional modules of an ASV system~\cite{spoof_review}. But this may not be feasible always as it requires access to various modules of the system. Another way to deceive a system is to craft so-called \emph{adversarial examples}~\cite{adv_samples_paper}. They are novel inputs crafted with some knowledge of the attacked system. Adversarial attacks have received a lot of attention across different classification tasks (especially within image processing) ~\cite{adversarial_learning}, but comparatively less in the speech field. While adversarial attacks and their defenses can be motivated from security improvement against `hackers', another viewpoint is general robustness improvement. Modern ASV systems are robust to many perturbations, but many spoofing countermeasures lack this property.


\begin{figure}[t!]
\begin{center}
\includegraphics[width=0.47\textwidth]{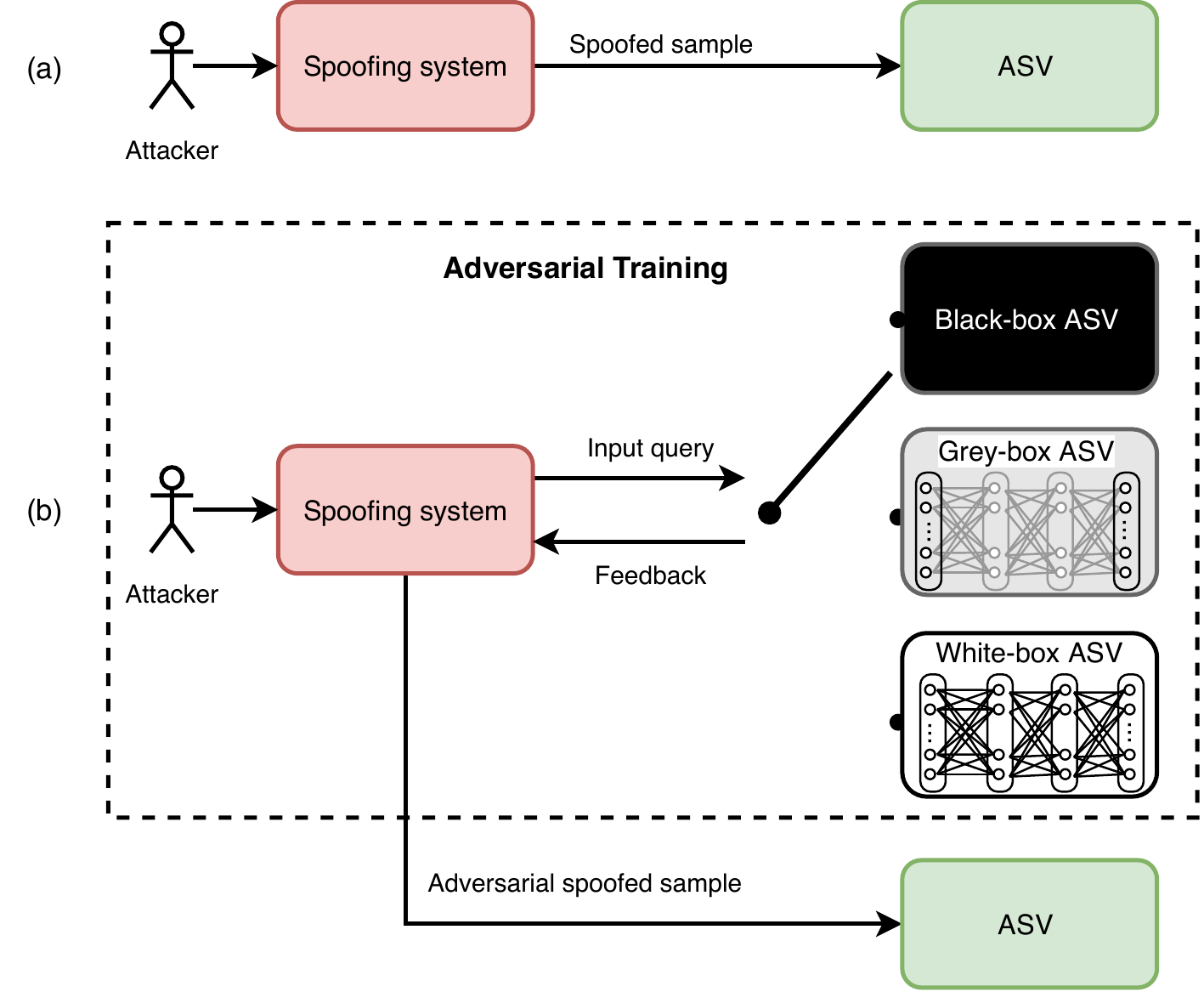}
\end{center}
\vspace{-3mm}
\caption{Spoofing from attacker's perspective (a) non-proactive attacks (b) adversarial attacks: using black-box, grey-box and white-box ASV.}
\label{attacker_fig}
\vspace{-6mm}
\end{figure}

Figure~\ref{attacker_fig} illustrates both non-proactive and  adversarial attacks from the perspective of the the attacker. In the latter case, the attacker leverages from information of the attacked ASV system to generate spoofed samples. 
The attacker can use the knowledge of either the attacked ASV or \emph{another} similar ASV to generate adversarial samples. The former is more effective (but potentially less realistic). Adversarial attacks can be broadly divided into \textbf{black-}, \textbf{grey-} and \textbf{white-box} attacks~\cite{adverserial_attacks_Yuan}. In the first case, the attacker 
has access only to the system output (speaker similarity score or hard accept/reject decision) to guide crafting of new inputs\cite{blackbox_adversary}. The grey-box attacks are a step further, where the attacker has some information such as features of the speakers and their implementation, but not their statistical models~\cite{grey_box_vivek}. Finally, the white-box attacks pose the greatest threat as the attackers have full knowledge of the system under attack~\cite{adverserial_attacks_Yuan}. Recent studies using adversarial attacks on various applications have demonstrated their threat to fool the system behavior~\cite{adv_samples_paper,Explain_adversary,physical_world_attack}. 

Studies specifically focused on adversarial attacks on ASV have come up only very recently~\cite{Fooling_ASV,VESTMAN_mimic_ASV,whitebox_SSW19,adversary_CM,ICASSP2020_sub} and some of this work has revealed new, potential threats. We present an overview 
of these studies. Non-proactive spoofing attacks 
are also briefly discussed for a broader context. In general, spoofing attacks can be performed either on the core ASV system, spoofing countermeasures or both. 
We group the studies on these grounds. 
Further, we discuss the results of different such attacks and possible emerging defense strategies against these attacks.




\section{Spoofing with Non-proactive Attacks}
\label{secii}

This section presents a review on traditional, or \textbf{non-proactive} spoofing attacks that use limited prior knowledge of the attacked ASV system. These attacks can be broadly divided into four categories, impersonation, replay, VC and TTS~\cite{spoof_review}. Impersonation is 
commonly referred as mimicry, where the attacker attempts to mimic the voice characteristics of the target speaker. Replay attacks are executed by replaying the previously recorded speech of the target speaker. Finally, VC and TTS attacks 
aim at modifying source speaker identity to that of a target speaker, and to produce text in a given target speaker's voice, respectively.

Fundamentally, crafting non-proactive 
attacks lacks a direct optimization target 
related to the attacked ASV system (such as false acceptance rate). Rather, such attacks represent ideas or technology originally designed with completely different aims and purposes in mind; they are taken \emph{as-is} to execute stress tests on ASV systems. For instance, mimicry takes place in acting and stand-up comedy without any reference to ASV systems. Similarly, VC and TTS technology researcher may not consider themselves as developers of `ASV attack technology' (any more than knife or gun manufacturers may consider themselves as developers of `murder technology'). Finally, speech recorders and loudspeakers (used in mounting of replay attacks) is technology intended to reproduce recorded or transmitted speech, music, or any other audio to a human listener as faithfully as possible. The fact that TTS, VC and replay attacks \emph{do} compromise the security of ASV systems is a lucky side-product\footnote{Implied by \emph{desirable} properties of the original technology, such as accurate reproduction of stored audio or target speaker voice timbre.} rather than the original aim. 
In the following subsections, we group the non-proactive attacks into three categories based on the their target to attack ASV with or without spoofing countermeasures.

\subsection{Attacks on ASV}

Different kinds of non-proactive attacks are investigated on ASV systems without spoofing countermeasures to showcase the impact of such attacks. Concerning mimicry attacks, impersonators tend to make vocal caricatures of their target speakers by mimicking high-level speaker cues such as prosody, accent, pronunciation and lexicon, more than the low-level spectral cues used by ASV systems. As a result, impersonation is not a consistent approach to attack ASV~\cite{mimic_vul,mimic_Tomi_journal}. 

In contrast to the mimicry attacks, the attacks generated by a VC or TTS system are optimized for both speaker similarity and quality. The aim of the former, a relevant concern for ASV, is to generate or modify speech so that it sounds as if spoken by a given target. This is often done by empirically minimizing a spectral distance measure~\cite{VC_ML,high_voice_morph} between the synthesized (or modified) and target speech, as a proxy of time-consuming perceptual experiments. Even if simple spectral distance measures have only a weak connection to speaker similarity computations implemented in ASV systems, several studies indicate that ASV systems are nonetheless vulnerable to these attacks~\cite{atrif_vc,TomiICASSP12,yamagishi2012,Tomi_ICASSP17}.
Further, modern VC and TTS systems are not tailored for a fixed set of speakers. High-quality target speaker voice can be generated by adapting an average voice model trained with multi-speakers' data towards the desired target 
~\cite{Tian2018} or by conditioning the model using a global (utterance-level) speaker variable~\cite{Hierarchical_SS}. 
These speaker-conditioning variables are similar (or even same) as \emph{speaker embeddings} used in ASV systems. These developments have made ASV and TTS/VC technologies closer to each other, imposing imminent threat to ASV systems.


ASV systems are also vulnerable to replay attacks which use pre-recorded speech samples of the target speaker~\cite{Li2016_spoof_TD}. As replayed samples contain strong traits of the target speaker, 
they pose a critical threat on any unprotected ASV system, most notably text-independent ASV and text-dependent ASV without protection against wrong passphrase. For ASV systems protected against wrongly spoken passphrase, the replay attacks require the pre-recorded samples of the same spoken content and are unflexible. On the other hand, the attacks derived using VC and TTS systems can be performed by only knowing the lexical information of the target speaker in a such scenario.  



\subsection{Attacks on Spoofing Countermeasures}

Spoofing countermeasures are introduced to the ASV systems to protect them from various attacks. An attacker may also try to attack only the spoofing countermeasures with non-proactive attacks that are not easily detectable. The studies on the first edition of ASVspoof challenge indicated that a \emph{unit-selection} based attack, produced by concatenation of time-domain waveform samples confused many of the spoofing countermeasures~\cite{ASVspoof_journal}. On the other hand,  attacks synthesized by using vocoders showed less threat to the spoofing countermeasures~\cite{ASVspoof_journal}.
A further study
~\cite{Kinnunen2018_vc} on the second edition of \emph{voice conversion challenge} \cite{Lorenzo-Trueba2018} suggested that modern waveform filtering based samples might be comparably harder to detect than traditional vocoded samples. 

For replay attack countermeasures, the replay configuration plays a key role. An analysis presented in~\cite{ASVspoofV2} suggested that replay speech generated with high quality recording and playback devices in clean environment can be particularly difficult to detect. This applies to \emph{any} kind of countermeasure as the artifacts distinguishing replay speech from the bonafide speech is minimal in such a scenario; whenever the replayed speech becomes digitally indistinguishable from bonafide speech 
no (low-level) countermeasure will be able to detect it.

\subsection{Attacks on ASV with Spoofing Countermeasures}

We have discussed attacks on ASV and countermeasures separately, but they could also be performed on combined systems consisting of ASV and countermeasures. There are no extensive studies on this direction.
Although~\cite{jointAVspoof} suggests that ASV with a spoofing countermeasure in combination might be less vulnerable to the attacks, the 
severity of attacks on ASV with spoofing countermeasure depends on the nature of their combination approach, which deserves further research.


\vspace{-2mm}
\section{Spoofing with Adversarial Attacks}
\label{seciii}

We now turn our attention to proactive, or
adversarial attacks, 
which have been explored for the case of TTS, VC and impersonation attacks. As far as the authors are aware of, replay attacks have not yet been investigated in an adversarial context.


\subsection{Attacks on ASV}


Optimizing input signals with partial or full knowledge of the attacked system is not a new concept in itself. For instance, \emph{artificial signals} (that may bear no resemblance to human speech) have been successfully used to attack ASV \cite{SPOOF2012evans}. A key difference in the adversarial attacks, however, is that the new signals are required to remain unnoticeable to human eye or ear --- being perceptually indistinguishable from natural signals. A study~\cite{Fooling_ASV} later generated adversarial samples by adding a perceptually indistinguishable structured noise to the original test examples for attacking an end-to-end ASV system.
Adversarial training uses the so-called \emph{fast gradient sign method} (FGSM)~\cite{Explain_adversary} with 
white-box and black-box attacks in a cross-corpora and cross-feature setting considering the same ASV. The studies demonstrated the ability of the adversarial attacks to deceive ASV systems. 
Another recent study ~\cite{adverserial_ASV_MengICASSP2020} also used FGSM 
to perform white-box and black-box attacks. It extended studies on adversarial \emph{transferability} from one ASV to attack another ASV system ~\cite{adverserial_ASV_MengICASSP2020}. 


Another adversarial attack against ASV, `FakeBob', is addressed in~\cite{WhoIRealBob}. This study uses black-box attacks by adding small perturbation to generate adversarial samples too, but considered different cases for practical scenario. These include studies with various ASV architectures (including commercial systems), transferability of attacks, practicality of over-the-air through replay and imperceptibility based on human perception. A further study, explored the real-time nature and feasibility of adversarial attacks replaying over-the-air by modeling room impulse response (RIR) during adversarial training~\cite{Adversarial_Attacks_ASV,RealTime_adverserial}.

The authors of~\cite{adversary_opt2019} investigated the effect of \emph{dictionary attacks} on ASV. This kind of attack allows  targeting large speaker population without having specific knowledge of individuals or their speech models~\cite{MasterPrint}. They selected a set of non-target trials that have high false acceptance in a population for an ASV system. Given such a trial and the training utterances of the speaker population, a time-domain waveform, \emph{master voice}, is learned by adding adversarial perturbations to maximize the spectrogram similarity. The time-domain waveform is generated by spectrogram inversion once the similarity exceeds a threshold to have a close match to a number of speakers in the population. The adversarial optimization of dictionary attacks were found to be imperative for deceiving ASV systems. 

A verification-to-synthesis attack using white-box ASV is carried out in~\cite{whitebox_SSW19}. In this adversarial attack, a VC system is trained using white-box ASV model without target speaker training data (unlike traditional VC systems). As the trained network may distort the phonetic properties of the input voice, an automatic speech recognition model is also included as part of optimization to regulate loss of phonemic information.
The output voice thus produced is not only able to deceive the ASV system, but also maintains the perceptual quality. 

The authors of~\cite{ICASSP2020_sub} studied black-box attacks on ASV through \emph{feedback-controlled} VC framework. The authors treat the ASV system as a black-box with access to its detection score only. This score is taken as a feedback to train the VC system. The objective function for training the feedback-controlled VC is jointly optimized with the feedback ASV score. 
The results indicated that black-box attacks can degrade ASV performance. Additionally, listening experiments suggested that adversarial examples are indistinguishable from the VC examples generated without ASV feedback.

The above studies typically assume that the system accessed or queried by the attacker is the same the as the attacker finally wishes to attack. In contrast to this assumption, there are also studies that assume that the attacker \emph{cannot} access the attacked system itself, but another ASV system, used as a proxy of the attacked one.
The authors of~\cite{VESTMAN_mimic_ASV} consider mimicry attacks where they find the closest target speaker for given attacker using a proxy ASV system. However, when asked to mimic their selected target speakers, the attackers did not manage to increase the detection score. The study indicated that mimicry, even when assisted by ASV-based, may not fool ASV. 
But an ASV system can be definitely used to assist the attacks on another ASV system.

\subsection{Attacks on Spoofing Countermeasures}

Adversarial attacks solely on spoofing countermeasures have received less attention. The authors of~\cite{SPSS_GAN_2017} proposed an adversarial training method for statistical parametric speech synthesis, where the loss function for training is modified by adding a weighted loss using an anti-spoofing system. As the loss function minimizes generation error as well as makes the distribution of synthetic speech close to that of natural speech, it is also able to deceive the anti-spoofing system apart from producing an improved speech quality. This work is extended for a generative adversarial network based synthetic speech generation framework, which also proved to be effective to increase the spoofing rate~\cite{SPSS_GAN}.  

A recent work in~\cite{adversary_CM} conducts white-box and black-box attacks on spoofing countermeasures. The authors consider one of the strong anti-spoofing system based on \emph{light convolutional neural network} (LCNN)~\cite{STC_ASVspoof2019} to carry out the adversarial attacks with the FGSM and the projected gradient descent methods. The studies conducted with both white-box and black-box attacks indicate that the well performing spoofing countermeasures can be fooled by generating adversarial samples. Further, listening test revealed that the adversarial samples are indistinguishable from non-proactive samples. 

\subsection{Attacks on ASV with Spoofing Countermeasures}

Adversarial attacks could also be carried out on ASV with spoofing countermeasure by leveraging from any prior information the attack has about either system.
As far as the authors are aware, there is currently no (reported) research on this direction. However, as many real-world systems combine ASV and countermeasures, future work should address attacks (both non-proactive and proactive) against combined system.

\begin{table*} [t!]
\caption{\label{table_discussion} {ASV and spoofing countermeasure (CM) performance before (B), after (A) adversarial attacks, and post defense (D) applied (if any) in different metrics, accuracy (ACC), equal error rate (EER), attack success rate (ASR), spoofing rate (SR) and score comparison.}}
\vspace{-2mm}
\centerline{
\resizebox{17.1cm}{!}{
\begin{tabular}{|c|c|c|c|c|c|}
\hline
{\bf Adversarial Attack/ Defense} &{\bf Attack Type} & {\bf ASV/CM System} &{\bf Corpus} & {\bf Performance (B/A/D)  } &{\bf Metric}\\
\hline
\hline
Adding perturbation~\cite{Fooling_ASV}  & White/black-box & End-to-end & YOHO, NTIMIT & 87.50/25.75/- (white-box) & ACC  (\%)\\
Adding perturbation~\cite{adverserial_ASV_MengICASSP2020}  & White/black-box & x-vector, i-vector  & VoxCeleb1 & 7.20/8.83/- (black-box, i-vector) & EER  (\%)\\
Adding perturbation~\cite{WhoIRealBob}  & Black-box & i-vector, GMM-UBM & LibriSpeech & -/70/- (i-vector) & ASR  (\%)\\
Adding perturbation with RIR~\cite{Adversarial_Attacks_ASV} & White-box & x-vector & VCTK & 10/50/- & ASR  (\%)\\
Adding perturbation with RIR~\cite{RealTime_adverserial} & White-box & x-vector & VCTK & 1.33/90.19/- & ASR  (\%)\\
Dictionary attack~\cite{adversary_opt2019} & White-box & VGGVox & VoxCeleb2 & -/20 (female), 10 (male)/- & SR  (\%)\\
VC with feedback loss~\cite{whitebox_SSW19} & White-box & d-vector & Japanese data & NA & Scores\\
Feedback-controlled VC~\cite{ICASSP2020_sub} & Black-box &i-vector & ASVspoof 2019 & 29.25/30.73/- & EER  (\%)\\
ASV assisted mimicry~\cite{VESTMAN_mimic_ASV} & Black-box & x-vector, i-vector  & VoxCeleb, self-collected & NA & Scores \\
TTS with feedback loss\cite{SPSS_GAN_2017,SPSS_GAN} & White-box & DNN & ATR Japanese & NA & SR plot\\
Adding noise~\cite{adversary_CM}  & White/black-box & LCNN, SENet & ASVspoof 2019 & 3.87/4.69/- (white-box, LCNN) & EER  (\%)\\
\hline\hline
Adding perturbation/  &\multirow{ 2}{*}{White-box} & \multirow{ 2}{*}{End-to-end} & \multirow{ 2}{*}{TIMIT} & {4.87/11.89/8.31 } & \multirow{ 2}{*}{EER  (\%)}\\
FGSM-REG, LDS-REG~\cite{Wang2019} & & & & (FGSM-REG) & \\
\hline
Adding perturbation/spatial  & \multirow{ 2}{*}{White-box} & \multirow{ 2}{*}{SENet, VGG} & \multirow{ 2}{*}{ASVspoof 2019} & {99.97/48.32/93.76} &
\multirow{ 2}{*}{ACC  (\%)}\\
 smoothing, adversarial training~\cite{adverserial_defensesICASSP20} & & & & (Adversarial training, SENet) & \\
\hline
\end{tabular}}
}
\vspace{-5mm}
\end{table*}

\vspace{-2mm}
\section{Defenses to Adversarial Attacks on ASV}
\label{seciv}

The spoofing conducted using adversarial attacks discussed in the previous section projects the weak spots of ASV. Although many countermeasures for non-proactive attacks are available, countermeasures for adversarial attacks need attention as well. In the field of machine learning, various defense mechanisms are employed to handle adversarial attacks~\cite{adversarial_attacks,adverserial_attacks_Yuan}. They can be categorized into \emph{passive} and \emph{proactive} defenses. The former aims to counter adversarial attacks without modifying the attacked system model. Proactive defenses, in turn, aim at training new models that are robust to adversarial samples. Motivated by such directions, there are some recent works that explore defense mechanisms against adversarial attacks on ASV. 

\emph{Adversarial regularization} is addressed in~\cite{Wang2019} to protect end-to-end ASV from adversarial attacks. The studies first generate adversarial samples by FGSM and \emph{local distributional smoothness} (LDS)~\cite{miyato2015distributional} method that are found to fool the ASV system. Therefore, the model is retrained with adversarial regularization as a defense mechanism. 
This mechanism aims at finding a worst spot around the current data point, and then optimize using this worst data point to derive a robust model~\cite{Wang2019}. The regularization is studied for both methods (FGSM-REG and LDS-REG) and is found to improve ASV performance against adversarial attacks.

Spoofing countermeasures also require defense mechanisms against adversarial attacks. A passive defense method namely, \emph{spatial smoothing}~\cite{spatial_smooth} and another proactive method namely, \emph{adversarial training} are studied to defend adversarial attacks for spoofing countermeasures~\cite{adverserial_defensesICASSP20}. The former is a simple method, where a slicing window moves over the power spectrum, then performs smoothing by use of filters such as median, mean and Gaussian, commonly used in the field of image processing. The general idea behind these simple noise suppression techniques is to suppress the impact of adversarial perturbations that are noise-like. The latter, in turn, leverages from adversarial samples at the training stage to improve robustness against attacks. Both methods are investigated on two spoofing countermeasures and are found as effective defense methods against the adversarial attacks.

\vspace{-2mm}
\section{Summary and Discussion}

Before concluding, we 
summarize the cited literature on adversarial attacks and their defenses
in Table~\ref{table_discussion}. 
Generally, the studies are diverse in terms of the adversarial attacks, attacked systems, datasets and metrics.
While these differences make it impossible to compare different studies, 
the available performance numbers within studies (when available) suggest that the adversarial attacks can severely degrade the ASV performance and that defenses are required for safeguarding systems from such attacks.  
Further, although the white-box attacks suggest a higher relative threat (as one might expect), black-box attacks might be more realistic; if the attacker already has full access to the system details, does he/she need to bother about generating spoofed samples? 


The defense mechanisms in~\cite{Wang2019,adverserial_defensesICASSP20} contributes to defend adversarial attacks as observed from Table~\ref{table_discussion}. However, these methods learn to resist \emph{particular kind of attack} in most cases. Therefore, such defense methods might be less effective when the attacker changes the settings of the attack~\cite{ICLR2017B}. 

To address such problem ensemble adversarial training is employed that generates a larger adversarial training examples by attacking several different models and then train the model by transferring the examples~\cite{adversarial_attacks}. This kind of defense mechanism might be more favorable from the outlook of practical systems, where the nature of adversarial attacks is always unknown. The challenges associated with unknown attacks has already been noted in the context of ASVspoof challenges. The evaluation data (provided without ground-truth to participants) have purposefully included some `surprises' --- attacks not included in the training data, and these have turned out difficult to detect.


We find the adversarial attacks that are proactive in nature, have a definite impact for knowing the weak spots of ASV systems as discussed throughout the paper. Nevertheless, the defense mechanisms to tackle such attacks are more imperative for improving system robustness in real-world scenario. 
This remains as an important direction for futuristic ASV systems.
\vspace{-2mm}
\section{Conclusions}
\label{conc}

The overview presented in this work shows that the proactive or adversarial attacks have a higher threat to ASV than the non-proactive attacks. However, they are less explored and the existing studies are dispersed across different dataset designs, different ways to evaluate various attacks and their defenses. Further, considering the practicality of adversarial attacks and their defenses, there is a need to have a common protocol, performance metric, and corpus for future research. The special session on {\emph{ The Attacker's Perspective on ASV}} in Interspeech 2020 organized by the authors is a small step towards this direction. 

\vspace{-2mm}
\section{Acknowledgements}

This research work is supported by Programmatic Grant No. A1687b0033 from the Singapore Government's Research, Innovation and Enterprise 2020 plan (Advanced Manufacturing and Engineering domain) and in part by the Academy of Finland (Proj. No. 309629 ``NOTCH: NOn-cooperaTive speaker CHaracterization'').

\vspace{-2mm}
\balance
\bibliographystyle{IEEEtran}


\end{document}